\documentclass[aip,amsmath,amssymb,reprint,jcp]{revtex4-1}
\usepackage{tikz}
\usepackage{graphicx}
\usepackage{dcolumn}
\usepackage{bm}
\usepackage{multirow}					
\usepackage{array}
\usepackage{footnote} 

\usepackage{booktabs}
\usepackage{float}
\usepackage{mathtools}
\usepackage{color}
\usepackage{hyperref}

\begin{document}

\title[]{Generating stable molecules using imitation and reinforcement learning}
\author{S{\o}ren Ager Meldgaard}
\affiliation{InterCat and Department of Physics and Astronomy, Aarhus University, Denmark.}
\author{Jonas K\"{o}hler}
\affiliation{Freie Universit\"{a}t Berlin, Department of Mathematics and Computer Science, Berlin, Germany}
\author{Henrik Lund Mortensen}
\affiliation{InterCat and Department of Physics and Astronomy, Aarhus University, Denmark.}
\author{Mads-Peter V. Christiansen}
\affiliation{InterCat and Department of Physics and Astronomy, Aarhus University, Denmark.}
\author{Frank No\'{e}}
\affiliation{Freie Universit\"{a}t Berlin, Department of Mathematics and Computer Science, Berlin, Germany}
\affiliation{Freie Universit\"{a}t Berlin, Department of Physics, Berlin, Germany}
\affiliation{Rice University, Department of Chemistry, Houston, TX, USA}
\author{Bj{\o}rk Hammer}
\affiliation{InterCat and Department of Physics and Astronomy, Aarhus University, Denmark.}
\email{hammer@phys.au.dk}

 \date{\today}

\begin{abstract}
Chemical space is routinely explored by machine learning methods to
discover interesting molecules, before time-consuming experimental
synthesizing is attempted. However, these methods often rely on a
graph representation, ignoring 3D information necessary for
determining the stability of the molecules. We propose a reinforcement
learning approach for generating molecules in cartesian coordinates
allowing for quantum chemical prediction of the stability. To improve
sample-efficiency we learn basic chemical rules from imitation
learning on the GDB-11 database to create an initial model applicable
for all stoichiometries. We then deploy multiple copies of
the model conditioned on a specific stoichiometry in a reinforcement
learning setting. The models correctly identify low energy molecules in
the database and produce novel isomers not found in the training
set. Finally, we apply the model to larger molecules to show how
reinforcement learning further refines the imitation learning model in
domains far from the training data.
\end{abstract}
\maketitle

\section{Introduction}
Discovering novel molecules or materials with desirable properties is
a challenging task because of the immense size of chemical compound
space. Further complicating the process is the costly and
time-consuming process of synthesizing and testing proposed
structures. Whereas this procedure historically was driven by a
trial-and-error process, the advance of computational quantum chemical
methods allows for initial screening to select promising molecules for
experimental testing. While the computational resource growth provided
a significant speed-up in processing molecules an exhaustive search of
molecular compound space is still infeasible. Instead an automated
search for interesting candidates is desired. Examples of such search
methods include evolutionary algorithms \cite{ga1,ga_mol},
basin-hopping \cite{basinhopping} and particle swarm optimization
\cite{PSO}. \\ More recently, machine learning (ML) enhanced versions
of aforementioned methods \cite{ga_ML,basinhopping_ML,PSO_ML} and
regression methods facilitating speed-up of virtual screening
\cite{VS1,VS2} has gained considerable interest by making researchers
able to quickly identify attractive candidates for experimental
testing. An added benefit of virtual screening is the creation of
numerous databases containing structures with computed chemical and
physical properties \cite{ZINC,GDB17,QM9,2D,catalysis} leading to
generative models for discovery of novel molecules and materials
\cite{GM0,GM1,GM2,GM3,esm2,GM4,SA3D,GM5,GM6,GM7,GM8,PVAE}. Unlike
virtual screening where candidates are selected among the structures
in the database, generative models have shown a remarkable ability to
leverage the database to produce new structures with desirable
properties. A notable limitation is a large database and no
feedback-loop to improve the generated molecules beyond what is
learned from the database. To remedy this, reinforcement learning (RL)
methods have started to become a competitive alternative
\cite{chemRL0,chemRL1,chemRL2,chemRL3,chemRL4,chemRL5,chemRL6} to
methods relying on existing databases. RL involves a model that
produces molecules and obtains properties for these molecules, by some
external means other than a database. This provides the basis for the
model to learn from the molecules it produces. Whereas generative
models rely on databases for pretraining, RL is usually done without
any prior knowledge leading to initial inefficiency as basic chemical
and physical rules are learned. Instead a reward must be defined
though, so it requires a setting where a meaningful reward function is
available, such as an energy or other things to be optimized. \\ In
this work we built upon a previous reinforcement learning algorithm
called Atomistic Structure Learning Algorithm (ASLA) \cite{asla} by
incorporating databases into molecular RL to improve sample efficiency
while simultaneously allowing the ML model to learn beyond the
knowledge contained in the database. First, a general purpose model is
trained using a database to create an model applicable for all
stoichiometries. Then, a copy of the general purpose model is created
for each stoichiometry of interest. These models are then further
refined in a RL setting in a search for low energy isomers for the
given stoichiometry. Specifically, we utilize a very small subset of
the GDB-11 \cite{GDB11} database that consists of small organic
molecules satisfying simple chemical stability rules and contains up
to 11 atoms of C, N, O and F. We demonstrate that our RL model is able
to both replicate structures in the database as well as producing
novel low energy structures. By focusing on low energy structures, we
believe we avoid the general pitfall of generative models that may
produce a large degree of unsynthesizable molecules, as e.g. shown in
the work of Gao and Connor \cite{gao}. Unlike previous approaches
using SMILES \cite{SMILES} or graphs, we operate directly in cartesian
coordinates thus allowing for optimization of the potential energy
hence easily biasing the search towards thermodynamically stable
structures. To improve sample-efficiency we take a model-based
approach where, similar to Ref.\ \onlinecite{asla2}, the potential
energy surface is modeled by a neural network allowing for approximate
but cheap optimization. Finally, we introduce an architecture based on
SchNet\cite{schnet} with a self-attention mechanism \cite{attention}
allowing the RL model to account for long-range effects. The paper is
constructed as follows: first the RL theory is outlined, followed by a
description of the neural network architecture. Then the database
pretraining and RL phase are described before demonstrating the method
on a subset of GDB-11. Finally, we apply the method to larger
molecules outside the training distribution.

\section{Theory}
\begin{figure} \centering
\includegraphics[width=0.9\columnwidth]{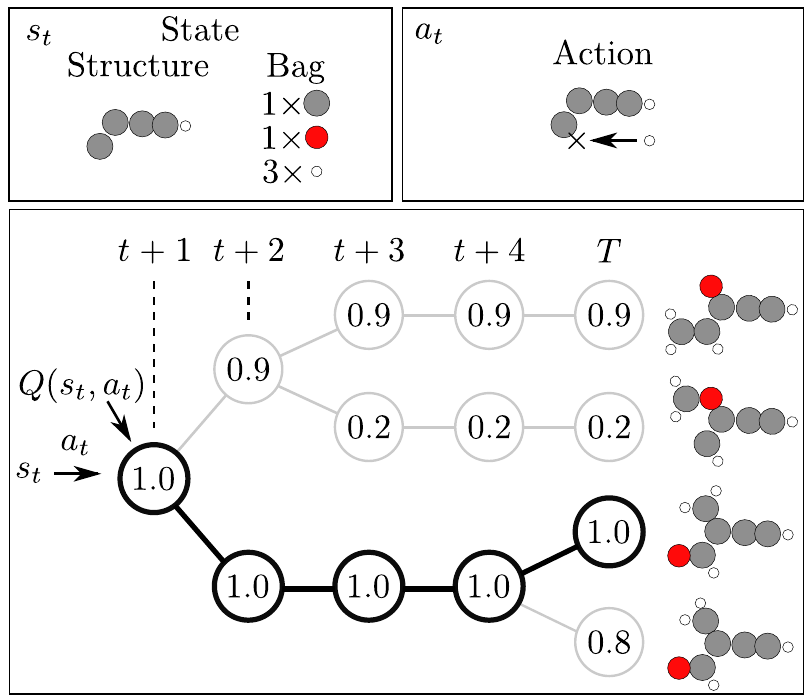}
  \caption{Given a state, $s_t$ and
an action $a_t$, the target $Q$-value is calculated by computing the
energy of finished structures involving the state-action pair. In the
above case a hydrogen atom is placed, which has led to four different
structures. The $Q$-value is updated towards the reward for the
lowest energy structure built.}
  \label{fig:1}
\end{figure}

The objective of RL is to solve a decision problem, i.e. formulate a
program that given an input can decide on the optimal
action. Specifically, we refer to this program as an agent which is
given a state, $s$, and must decide on an action, $a$. In the general
case the decision process involves multiple steps indexed with the
subscript $t$ while $T$ is used for the final step. In order to
solve the problem the agent must device a policy, $\pi$, which is a
probability distribution over the possible actions in a given
state. The optimal policy is the policy which maximizes the sum of
rewards, where $r(s)$ is the reward given in state $s$. The states,
actions and rewards are user-specified and describes the problem to be
solved. Following Ref.\ \onlinecite{chemRL6}, the state space is
defined as all (possibly partial) molecules along with a bag of atoms
not yet attached to the molecule. For the reward we seek to minimize
the potential energy, i.e.\ we set intermediate rewards to zero and
assign a final reward based on the potential energy of the
molecule.

\begin{align}
  r(s_t) = \begin{cases} \max\left(\frac{E_{\text{ref}} -
E(s_T)}{\Delta E} + 1, 0\right), & \mbox{if $t = T$} \\ 0, &
\mbox{else} \end{cases}
\end{align}
Here $E_{\text{ref}}$ is the lowest energy observed and $\Delta E =
10\text{eV}$ is the energy span from $E_{\text{ref}}$ where energy
differences are resolved. Note that $E_{\text{ref}}\le E(s_T)$ meaning
that $0\le r(s_t)\le 1$. We scale the reward to stabilize the training
as energies of produced structures may fluctuate substantially during
the RL phase. To maximize the expected sum of rewards we define the
$Q$-value,

\begin{align}
  Q_\pi(s,a) &= \mathbb{E}_{\pi} \left[\sum_{k = t + 1}^{T} r(s_{k})
\middle| s_t = s, a_t = a\right] \nonumber \\ &= \mathbb{E}_{\pi}
[r(s_T)| s_t = s, a_t = a],
  \label{eq:stateaction}
\end{align}
i.e.\ the expected final reward when taking action $a$ in
state $s$ and then following policy $\pi$. The goal of the agent is to
infer the optimal $Q$-value function
\begin{align}
  Q_*(s,a) = \max_{\pi} Q_\pi(s,a) \; \forall s, a,
\end{align}
thereby enabling the agent to perform the best action in every
state. To improve the $Q$-value function we parameterize it by a
neural network and update it as the agent collects new
experience. Specifically, we evaluate the $Q$-values on a voxelated
grid, which allows us to update the $Q$-value towards the highest
final reward observed by the agent for a specific state-action pair
(Fig.~\ref{fig:1}), which for a deterministic problem puts a lower
bound on the optimal $Q$-value \cite{brute}. Additionally, this
discretization allows for easy inference of the optimal atom placement
by simply finding the voxel which maximizes the $Q$-value.

\section{Architecture}
\begin{figure*}[t] \centering
\includegraphics[width=0.95\textwidth]{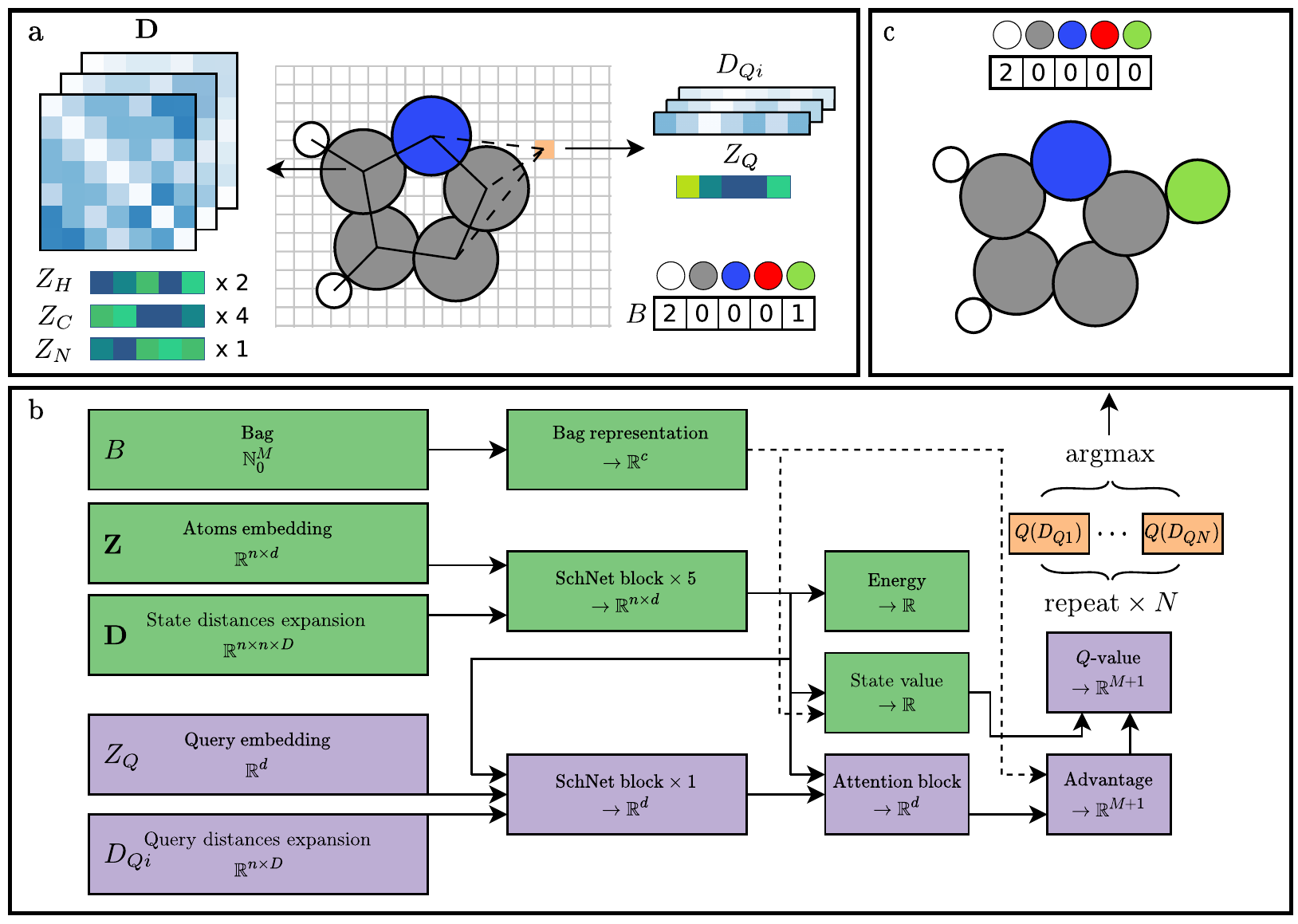}
  \caption{(a) The molecule is represented as a distance matrix $D$
(solid lines between atoms) in a Gaussian basis, embeddings for each
atom type $(Z_H, Z_C, Z_N)$ and a bag $B$. An action constitutes an
atom placed in one of the voxels. Each action is represented as a
distance to the atoms in the state $D_Q$ (dashed lines, not all
distances shown) and a query embedding $Z_Q$. (b) To predict $Q(s,a)$ the
network takes as input the state and action representation. The green
part of the network operates on the state and the purple section
includes the action. The action part is repeated for each voxel in the
state resulting in $N$ possible actions. (c) The resulting state after
the voxel with the highest $Q$-value is chosen. If the state is
terminal the structure is relaxed using the model potential.}
  \label{fig:2}
\end{figure*}

The state representation is utilized for calculating the energy
of a structure as well as a state value which is used for
calculating the $Q$-value, but is independent of the specific
action. For improved data efficiency the state and action
representation must satisfy symmetries in the Hamiltonian, i.e.\
translation, rotation and mirroring of the molecule as well as
permutation of identical atoms. To incorporate these properties in the
state representation we utilize pairwise distances (Fig.~\ref{fig:2}a,
$\mathbf{D}$), expanded in a Gaussian basis
\begin{equation}
  \mathbf{D}_{ijk} = e^{-\gamma (r_{ij} - \mu_k)^2} f_c(r_{ij})
\end{equation}
where $\gamma = 1$\AA$^{-2}$ and $r_{ij}$ is the distance between
atom $i$ and $j$. For $\mu_k$, we choose 20 values uniformly between 0
and a cutoff-radius $r_c = 5$\AA. The cutoff function
\begin{equation}
  f_c(r) = \begin{cases}
\frac{1}{2}\left(\cos(\frac{\pi r}{r_c}) + 1\right), & \mbox{if $r \le
r_c$} \\ 0, & \mbox{else} \end{cases}
\end{equation}
emphasizes the importance of the local neighborhood by decaying
$\mathbf{D}_{ijk}$ as a function of $r_{ij}$ and ensures a smooth
transition to zero at $r_c$. For the atoms, we utilize randomly
initialized trainable atom type embeddings (Fig.~\ref{fig:2}a, $Z_H,
Z_C, Z_N$). Finally, the bag is represented by the number of remaining
atoms (Fig.~\ref{fig:2}a, $B$).\\ The action is represented by
distances to atoms already placed (Fig.~\ref{fig:2}a, $D_Q$) as well
as a special query type embedding (Fig.~\ref{fig:2}a, $Z_Q$). When
calculating $Q$-values the query atom is placed at various voxels
allowing for inference of $Q$-values at possible positions of the next
atom to be placed. In this way, all the encoded distances are
easily converted to a 3D structure.

The state representation (Fig.~\ref{fig:2}b, green) is updated
using SchNet-blocks. In the following we will use superscripts to
index the blocks, if multiple blocks follow each other. Fully connected
layers are given as $W_i$ where the full details for each
fully connected layer is given in the supplementary information. Each
block operates on the distance and atom type embeddings using
continuous-filter convolutions with skip connections as introduced in
SchNet:
\begin{align}
  \tilde{Z}^{l+1}_i = Z_i^l + \sum_{j \neq i} Z^l_j \circ
W_d^l(\mathbf{D}_{ij}),
\end{align}
where $\circ$ is elementwise-multiplication and $Z_i \in \mathbb{R}^d$
are the atom embeddings. The convolution is followed by an atomwise
fully connected layer with a skip connection:
\begin{align}
  Z^{l+1}_i = \tilde{Z}^{l+1}_i +W_a^l(\tilde{Z}^{l+1}_i),
\end{align}
which is passed to the next SchNet block. By chaining several blocks,
information from distant atoms are passed to each atom
embedding. After $L = 5$ layers, the atomwise representations,
$Z^L_i$, are used for predicting the energy $(E)$ of the structure as
as combination of local contributions
\begin{align}
  E = \sum_i W_E(Z^L_i)
\end{align}
Additionally, we compute a state value ($SV$).
\begin{align}
SV = W_{SV} \left(W_B(B) \oplus \sum_i Z^L_i\right),
\end{align}
where $\oplus$ is concatenation. The state value is an indication of
whether high or low $Q$-values are expected from the current state, as
will be evident in the next section. Unlike the energy, the state
value must consider remaining atoms in the bag when calculating if the
current state can progress into a valid molecule. \\ 

Having covered the part of the architecture that deals with the state,
we now turn to the action part given in Fig.~\ref{fig:2}b, purple). The
query type representation is updated using a
single SchNet-block. Furthermore, a multi-head self-attention block
over all the atom representations follows. In contrast to local models
where information from distant atoms must propagate through several
blocks, self-attention allows for immediate global information
flow. In the case of predicting $Q$-values for unfinished structures
this is especially relevant, as the model must be aware of possible
dangling bonds in one end of a molecule while predicting $Q$-values in
the other end. For $h = 8$ heads, we calculate a query, key and value as
\begin{align}
  q^l_i = Q^l(Z_i), \; k^l_i = K^l(Z_i), \; v^l_i = V^l(Z_i),
\end{align}
where the superscript index the head and $Q^l, K^l, V^l$
are linear projections to a subspace of size $ d_h = d / h$, where
$Z_i \in \mathbb{R}^d$. From this, a dot-product attention score is
calculated
\begin{align}
  \alpha^l_{ij} = \frac{\exp{(q^l_i \cdot k^l_j / \sqrt{d_h})}}{\sum_j
\exp{(q^l_i \cdot k_j / \sqrt{d_h})}},
\end{align}
which indicates the importance of latent representation
$v^l_j$ to describe $v^l_i$. From these scores a new latent
representation of $Z_Q$ is created as
\begin{align}
  Z_Q^l = \sum_j \alpha^l_{Qj} v_j^l
\end{align}
which is finally concatenated and processed by a fully connected layer with
a skip connection and layer normalization \cite{layernorm}
\begin{align} \tilde{Z}_Q = \text{Layer-norm}[Z_Q + W_c(Z_Q^1
\oplus \dots \oplus Z_Q^h)]
\end{align}
The representation of the query
atom is then used for calculating the $Q$-value
\begin{align}
  A &= W_A(\tilde{Z}_Q) \\
  Q &= \text{Softmax}(SV + A),
\end{align}
where $A \in \mathbb{R}^{M + 1}$ , $M$ is the number of
atom types, and $SV \in \mathbb{R}$ is added to the first $M$ entries
of $A$. Entry $i$ in $Q$ specify the $Q$-value for atom type $i$ and
the $M + 1'\text{th}$ entry allows for low $Q$-values for all atom
types. $SV$ enables the network to lower the $Q$-value for all
possible actions in a given state, instead of independently learning
that all actions leads to high-energy structures. By evaluating the
$Q$-value at multiple voxels, the agent follows the greedy policy of
picking the action with the highest $Q$-value, leading to the
structure in Fig.~\ref{fig:2}c.

\section{Method}
The algorithm consists of two phases. In the first phase we use a
simple form of imitation learning (IL) called behavioral cloning,
where the agent learns to build the structures in the database as well
as predicting their energies and forces. This agent serves as a
starting point for all agents employed in the next phase. \\ In the
second phase, a RL process is started where stoichiometric-specific
agents refines the IL model by constructing new molecules and querying
an energy calculator. Based on the received energies and forces, the
model updates the $Q$-values, energies and forces before constructing
a new molecule. As the agents receive feedback from the energy
calculator the molecules continue to improve.

\subsection{Imitation learning phase}
To create an initial model the GDB-11 database consisting of molecules
containing up to 11 heavy atoms (C,N,O,F) is utilized. Specifically,
we use all 1850 structures with 6 heavy atoms for supervised
pretraining. Using RDKIT \cite{RDKIT} SMILES are transformed into 3D
structures and a single point density functional theory (DFT)
calculation of the energy and forces is performed using GPAW
\cite{gpaw1,gpaw2} using a localized atomic basis set~\cite{lcao}. The
model is then trained using the following loss function
\begin{equation}
  l = \rho_E l_E + \rho_F l_F + \rho_Q l_Q,
\end{equation}
where we have
\begin{align}
  l_E &= ||E - \hat{E}||^2 \label{eq:E} \\
  l_F &= \frac{1}{3N} \sum_{i =1}^{3N}\text{Huber}(F_i, \hat{F}_i) \label{eq:F}\\
  l_Q &= - \sum_{i = 1}^{M + 1} p_i \log{\hat{Q}_i}, \label{eq:Q}
\end{align}
where the hat denotes values predicted by the network and $\rho_E =
0.1$, $\rho_F = 0.9$ and $\rho_Q = 1$ are empirically chosen to
balance the contributions to the total loss. $N$ is the number of
atoms in the structure, with $M$ different atom types. \\
The energy (eq.~\ref{eq:E}) is trained using a mean squared error (MSE) loss while
forces (eq.~\ref{eq:F}) are updated using a Huber loss given by
\begin{align} \text{Huber}(F, \hat{F}) = \begin{cases} \frac{1}{2}(F -
\hat{F})^2, & \mbox{if $|F - \hat{F}| < 1$} \\ |F - \hat{F}| -
\frac{1}{2}, & \mbox{else} \end{cases}
\end{align}
i.e.~a squared loss for small differences and a linear loss for large
errors. The Huber loss was more stable by suppressing
the effect of large force outliers which were occasionally present in
the structures built by the model. \\ 
Similarly, a cross entropy loss for the $Q$-values (eq.~\ref{eq:Q})
was found to be more stable than a MSE for the pretraining phase where
only $Q$-values of 0 or 1 are required. For the cross entropy loss
$\hat{Q}_i$ refers to the $Q$-value for atom type $i$ and we set $p_i
= 1$ for the atom type placed in the database and 0 for the other
types. Additionally, for each action present in the database we
generate five perturbed actions by randomly moving the selected atom
and setting $p_{i} = 1$ for the $M+1'$th entry in the $Q$-value
prediction, thereby enforcing $Q$-values for all atom types to be zero
for the randomly generated actions. As this generates an imbalance in
the training data the valid actions are weighted by a factor of
five. \\ 
The model is implemented in Pytorch \cite{pytorch} and trained using
the Adam \cite{adam} optimizer with an initial learning rate of $5
\cdot 10^{-3}$ and a batch size of 384. We use 90$\%$ of the
structures for training the remaining $10\%$ are used as a validation
set for decaying the learning rate by a factor of 2 when the
validation error plateaus for 30 epochs. Once the learning rate
decayed below $10^{-6}$ the validation loss stagnated, and the final
model is chosen for employment in the RL phase.

\subsection{Reinforcement learning phase}
In the RL phase we deploy the IL model for all 135 stoichiometries
present in the training set. A cubic cell of dimension [20\AA, 20\AA,
20\AA] and a $Q$-value grid resolution of 0.2{\AA} is chosen, with the
initial atom restricted to the center of the cell. We use a modified
$\varepsilon$-greedy strategy, to trade-off exploration and
exploitation. To satisfy two random moves per episode in expectation,
we choose a greedy action with probability 1-2/T and a random action
with probability 2/T, where $T$ is the number of atoms in the
structure. We choose $5\%$ of random actions uniformly random, while
the other $95\%$ are uniformly sampled from the top $5\%$ of
actions. Furthermore, all hydrogen atoms are masked until all other
atom types are placed which reduces the action space without excluding
any molecules. Due to the IL model an action among the highest
$Q$-values is frequently taken which ensures exploration while
simultaneously suppressing the large fraction of actions resulting in
completely invalid molecules.\\
To limit the action space, we only allow actions which fulfill 
\begin{align}
  c_1[r_{\text{cov}}(a) + r_{\text{cov}}(i)] < r_{ai} < c_2
[r_{\text{cov}}(a) + r_{\text{cov}}(i)],
\end{align}
for at least one atom $i$ already present in the structure. Here $c_1
= 0.75$, $c_2 = 1.25$ and $r_{\text{cov}}(i)$ is the covalent radius
of atom $i$. $r_{\text{cov}}(a)$ is the covalent radius of the atom
placed in action $a$ and $r_{ai}$ is the distance between the new atom
and a present atom $i$. That is, the new atom must be placed within
0.75 to 1.25 times the sum of the covalent radii of itself and an
already present atom. This restricts the model to building
non-fragmented structures for a better comparison with the
database. \\
The agent trained as detailed in the previous section is perfectly
capable of building a variety of different molecules given a
predetermined stoichiometry. Owing to the stochastic nature of the
modified $\varepsilon$-greedy policy employed sequential builds using
the same agent results in different structures being built. \\
We exploit that to construct a first ensemble of structures by tasking
the agent with building molecules 200 times. It does so according to
the pretrained $Q$-values and every completed structure is subject to
a relaxation in the model energy, eq. (8), followed by snapping the
molecule back on the grid. If the relaxation results in a
fragmentation of the molecule into several pieces, the original
structure is used instead. The initial exploration phase promotes
exploration of a large part of configuration space to avoid premature
convergence to a local minimum. Since the model is pretrained the
generated molecules will generally be of high quality. \\
Once the first 200 builds are completed, the search enters the RL
mode. Here the algorithm alternates between generating a structure and
updating the model using five mini-batches until a total of 600
additional molecules have been generated. For the $Q$-values the loss
function is changed to MSE, since the target is now given by
eq. (\ref{eq:stateaction}), i.e.\ it is no longer 0 or 1, but depends
on the potential energy of the molecule. By further improving the
$Q$-values and force-predictions the search is directed towards
low-energy regions of configuration space.

\subsection{GDB-11 benchmark}
\begin{figure}
  \centering
\includegraphics[width=0.9\columnwidth]{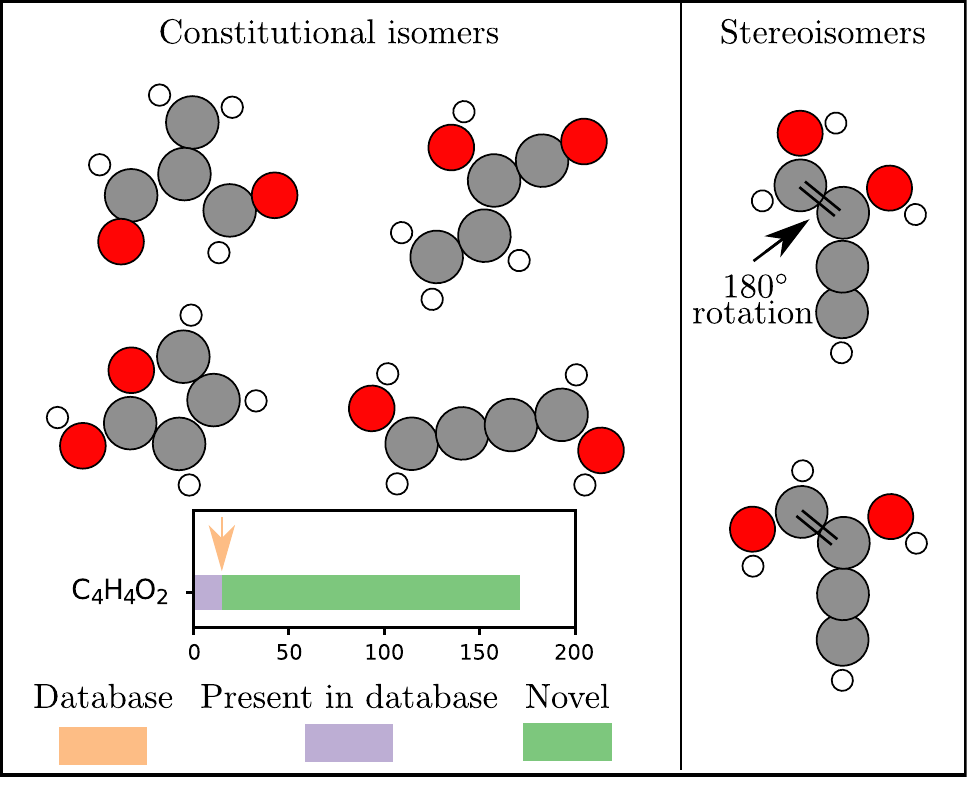}
  \caption{Subset of discovered constitutional isomers (left) and
stereoisomers (right) for C$_4$H$_4$O$_2$. Further shown is the number of all
unique constitutional isomers in the database (orange arrow) as well as
the molecules produced by ASLA divided into molecules present in the
database (purple bar) and novel molecules (green bar).}
  \label{fig:isomers}
\end{figure}

We start by testing the performance of the current framework when it
comes to identifying molecules with stoichiometries already present in
the database used for the training. The search is conducted as a set
of 64 independent restarts following the outlined protocol: using the
data-based-trained agent for 200 builds and applying RL for 600
subsequent builds. To facilitate the quantification of the amount of
new structures found, we resort to the use of SMILES, as an analytic
tool. We stress that SMILES are not used in any way by the agents as
this would limit the scope of these. Once the search is finished all
structures are post processed by taking five gradient steps using the
\textit{final} energy landscape. This is a computationally cheap step
that removes uncertainties from the original relaxation of the
molecules in the less reliable model energy landscape that had been
learned up to the point of the RL generating the particular molecule as
well as unfortunate grid snapping. These re-relaxed structures are
then converted to SMILES and all unique constitutional isomers are
found. To this end, we show in Fig.~\ref{fig:isomers} some of the
molecules being built when searching for C$_4$H$_4$O$_2$. As
illustrated, we find both new constitutional isomers and often several
stereoisomers for these. Looking across these 64 restarts, all of the
structures of this composition present in the database (orange arrow)
are found again (purple bar). However, also a considerable number of
molecules not present in the database are found (green bar). 

\begin{figure}[h]
  \centering
\includegraphics[width=\columnwidth]{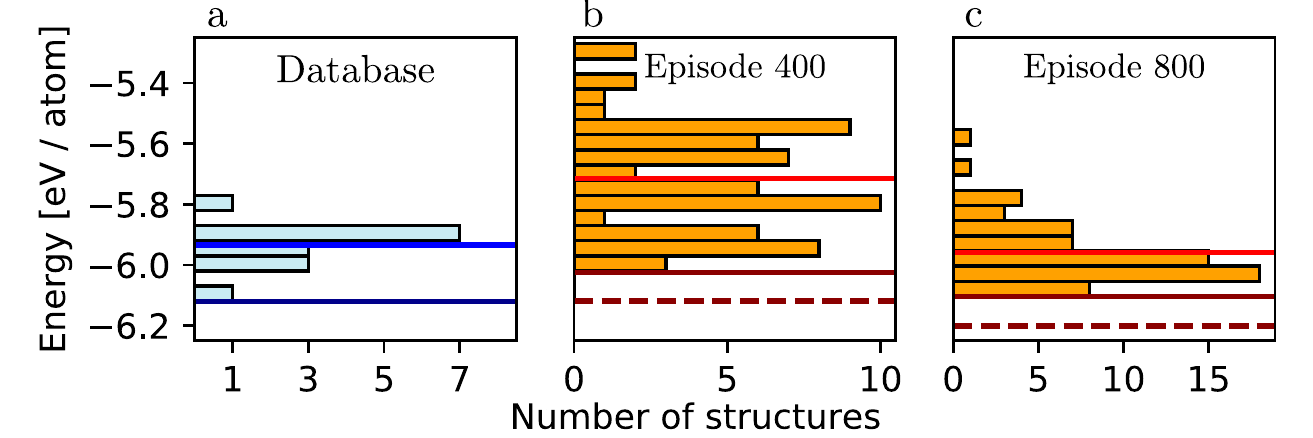}
  \caption{DFT energy distribution of the molecules. (a) Molecules in the
database. Blue line is the average energy and dark blue line is the
lowest energy isomer. (b) Orange: structures built in episode 400. The
red line indicates the mean energy of the structure and the dark red
shows the mean of the lowest energy structure in each run found before
episode 400. Finally, the dark red dashed line shows the lowest energy
isomer among the 64 parallel runs, found before episode 400. (c) Same
as (b), but now at 800 episodes. As the search progresses a shift
towards lower energies is observed.}
  \label{fig:C4H4O2_runs}
\end{figure}

To investigate whether the produced molecules are of low energy we
illustrate the evolution of the 64 agents from episode 400 to 800 in
Fig.~\ref{fig:C4H4O2_runs}. The histograms indicate the energy
distribution of the structures built in the given episode across the
64 agents. As the search progresses the distribution shifts towards
lower energies as the model begins converging, which is further
evident by the mean of the distribution (red line) shifting
down. Similarly, when we only consider the mean of the lowest energy
structure in each run (i.e. not necessarily built in episode 400 or
800), which is depicted using dark red we see a similar shift. In blue,
the average energy of the structures in the database, snapped to the
grid ASLA operates on, is given. Initially the database is better than
the average built of the agent due to imperfect fitting to the
database but is overtaken by the agent at episode 800. If one looks
only at the best structure in the database (dark blue) the average
lowest energy approaches this limit and is marginally beaten by one
of the isomers (dark red, dashed). Fully relaxing all unique isomers in
DFT indeed reveals that the lowest energy isomer produced by ASLA
coincides with the minimum structure in the database.

\begin{figure*}[t]
  \centering
\includegraphics[width=0.95\textwidth]{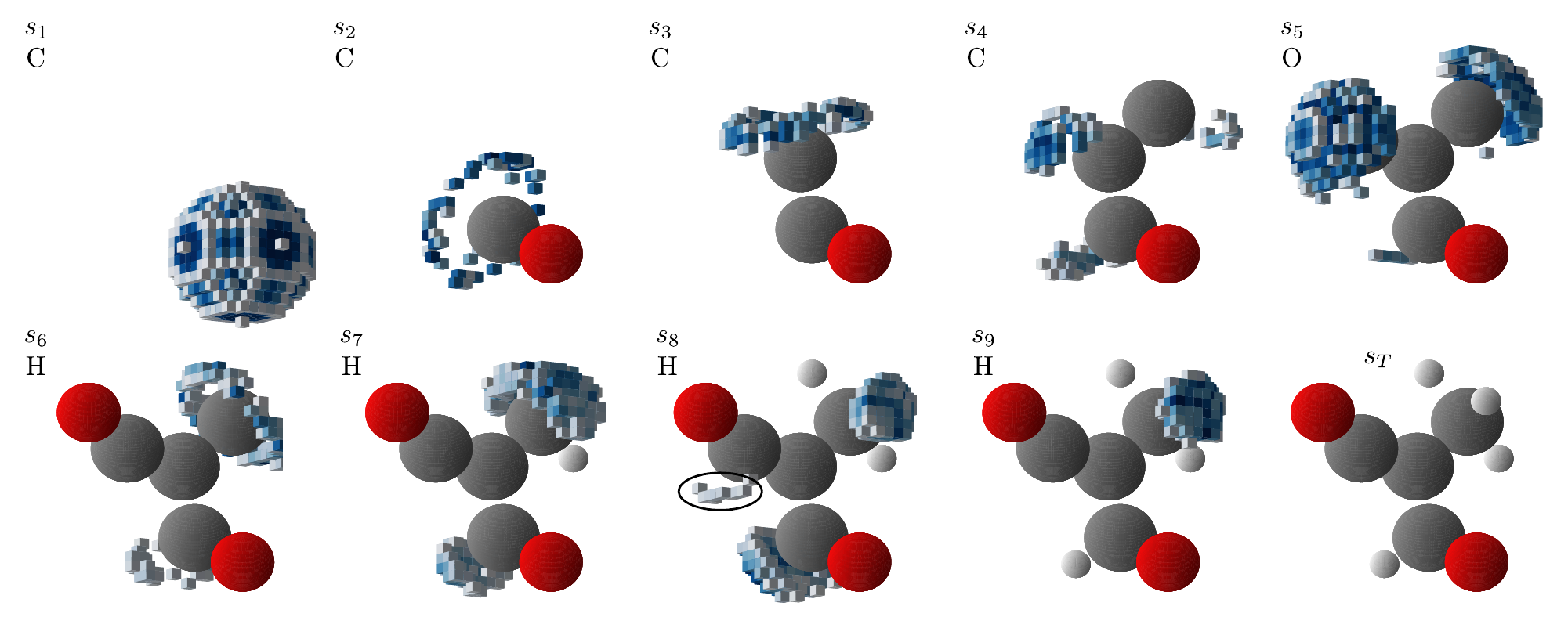}
  \caption{All $Q$-values above 0.7 are shown with darker blue
corresponding to higher $Q$-values. For each state we only show the
$Q$-values corresponding to the atom placed in the given step.}
  \label{fig:qvalues}
\end{figure*}

To observe the decision process for the RL algorithm we plot the
$Q$-values for one of the runs in figure \ref{fig:qvalues}. First of
all, the model has clearly learned a great deal about chemical bonding
and thus generally suppresses invalid actions, predicting only high
$Q$-values for sensible bonding sites. The agent is fully aware of
possible symmetries during the build. Starting with a single atom in
$s_1$, the $Q$-values have spherical symmetry. For the linear dimer in
$s_2$ the symmetry of the $Q$-values reduces to a ring, and for the
Y-shaped C$_3$O backbone in $s_5$, two symmetric 'caps' of high
$Q$-values develop. As evident in states $s_4$-$s_8$ where multiple
sites have high $Q$-values, the model decides between different
possible isomers when building the molecule. Bond lengths chosen early
on e.g.\ the C-C bond lengths when placing the C atoms from $s_2$ to
$s_5$ have an impact on later $Q$-values. The $Q$-values encircled by
a black ellipse in $s_8$ are thus smaller than the $Q$-value
that are chosen upon building $s_9$. In both cases, aldehyde (HOC-)
groups are formed, but at $C$ atoms involved in double or single C-C
bonds, respectively, the latter being the energetically
preferred. Placing H atoms at both sites would have been possible
causing tautomerization involving the methyl group, but again given
the initial choices for C-C bond lengths, this is no longer preferable
as reflected by the $Q$-values in $s_9$. Owing to the modified
$\varepsilon$-greedy policy, such alternative molecular builds are
occasionally made despite the C-C bond lengths, and the subsequent
relaxation in the on-the-fly learned model will provide
properly adjusted molecules for the RL training.\\

\begin{figure} \centering
\includegraphics[width=0.9\columnwidth]{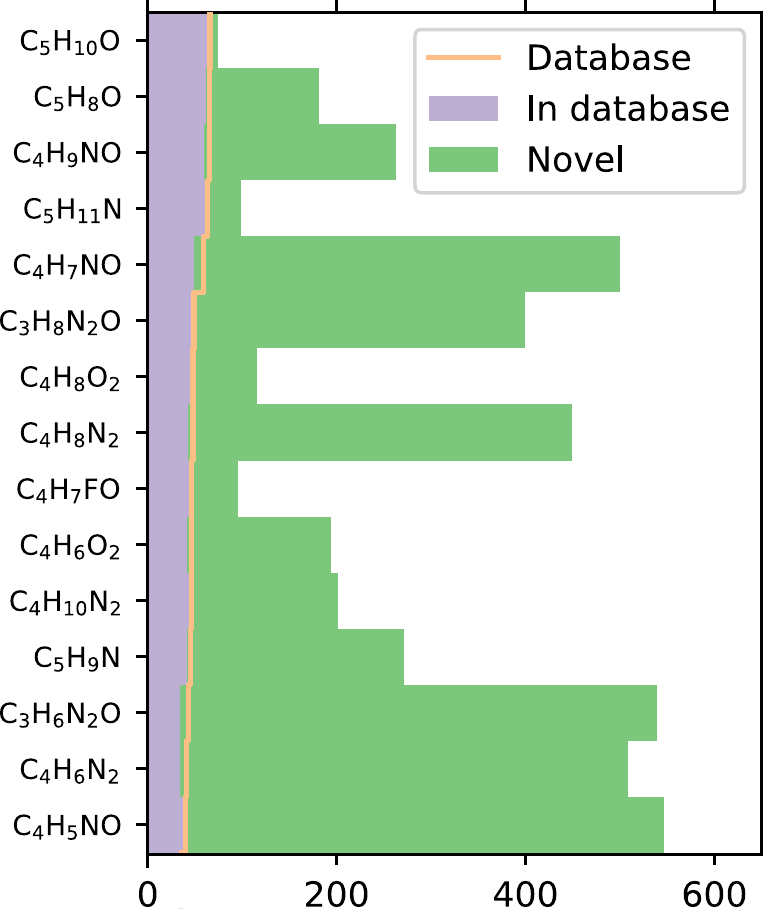}
  \caption{Molecules discovered. Orange line: number of constitutional
isomers in the database. Purple bars: molecules found that are also
present in the database. Green bars: novel molecules found.}
  \label{fig:4}
\end{figure}

In Fig.~\ref{fig:4} the total number of structures found after 800
episodes for the 15 stoichiometries with most isomers in
the database are shown. For a figure involving all 135
stoichiometries, see supplementary Fig.~\ref{fig:s1}. Similarly to
earlier, purple bars indicate the structures present in both the search
and the database, whereas green indicate novel structures. A majority
of the structures in the database (orange line) are found within the
800 episodes using 64 searches. In total 1769 of the 1850 molecules in
the database are found, i.e.\ slightly more than the 1665 structures
used for the training set. Accounting for both novel molecules and
structures rediscovered from the database, 20074 unique constitutional
isomers are generated resulting in a 10-fold increase of the
database. Especially for stoichiometries where multiple bond
combinations are possible, i.e.\ for stoichiometries containing a lot of
carbon, oxygen and nitrogen several new constitutional isomers are
discovered, whereas for hydrocarbons and C$_x$H$_y$F$_z$ where a lot
of single bonds are present the database covers a large fraction of
the isomers.

If the RL is neglected and the IL model is used for sampling
in all 800 episodes, 1649 of the structures in the database and 15709
novel molecules are generated for a total of 17358 unique
constitutional isomers. The IL model thus covers a large
fraction of the produced structures and turning on RL provides 15.6\%
more molecules. \\

\begin{figure} \centering
\includegraphics[width=0.9\columnwidth]{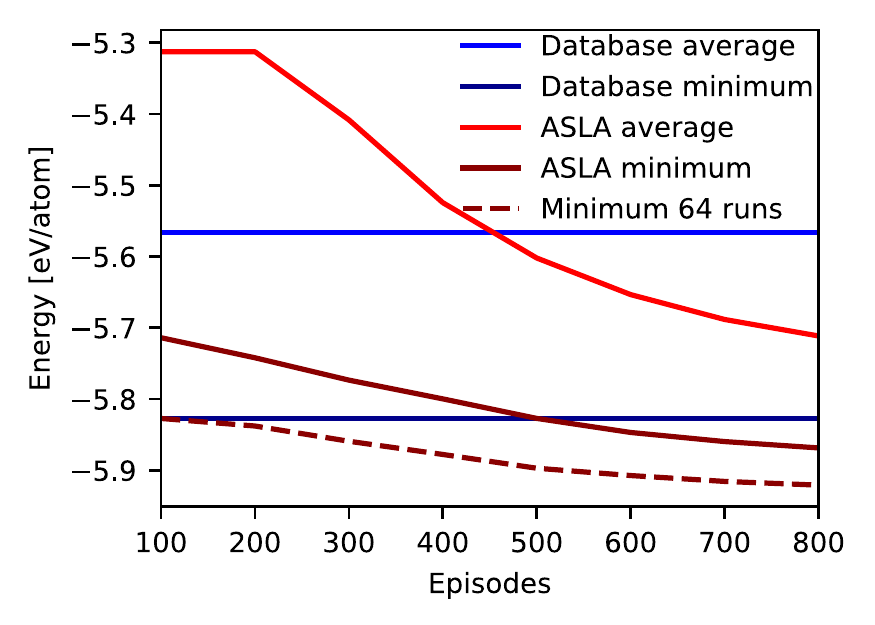}
  \caption{Blue: average energy of structures in the database. Dark
blue: average energy of lowest energy molecules in the database. Red: average energy
of structures built by ASLA. Dark red: average energy of lowest
energy molecules found by ASLA. Dashed dark red: lowest energy isomer among the 64 runs.}
  \label{fig:5}
\end{figure}

To illustrate the difference between the IL model and the RL
version, we plot the average atomization energy per atom of the
structures generated by ASLA throughout the 800 episodes
(fig.~\ref{fig:5}) and compare it to the energies of the structures in
the database when placed on the same 0.2{\AA} resolution grid. For
each ASLA curve the average is over the 64 independents run as earlier
and now also the 135 stoichiometries. As previously observed, the
model trained on the database starts out by producing higher energy
structures (red curve) than the average structure in the database
(blue curve), due to the imperfect fitting of the database. Similarly,
the average of the best structures (dark red) is worse than the lowest
energy isomer in the database (dark blue). At around 500 episodes both
the average and best structures produced by ASLA marginally
outperforms the database. At 800 episodes the best structure is
outperformed by $0.04$ eV/atom on average. If one only looks at the
best structure in each of the 64 runs (dark red, dashed) the database
is outperformed by $0.09$ eV/atom.

In order to check if this energy difference amounts to new low energy
structures the specific \emph{stereoisomers} in the database and the
structures generated by ASLA are compared. In that case 1742 of the
molecules in the database are found. When counting
\emph{constitutional} isomers as in Fig.~\ref{fig:4}, 1769 of the
structures in the database were found, which means that in 27 cases
the stereoisomer in the database is not found, but the constitutional
isomer is. In total, 29049 stereoisomers not present in the database
are found. Each stereoisomer in both the database and generated by
ASLA is then fully relaxed using DFT, which shows that for each
stoichiometry the lowest energy stereoisomer in the database is found
by ASLA for all but one stoichiometry and in 72 of 135 stoichiometries
a lower energy stereoisomer is found.

\subsection{Exploration beyond the database}
Having shown that the ASLA framework is capable of
reproducing and expanding molecular structures of compositions
already present in the database, we now move to explore its
performance when applied outside the realms of the
database. 

As an example, we investigate C$_9$H$_8$O$_4$, a
molecule with twice as many heavy elements as in the training set. As
before, 64 independent runs are started.

\begin{figure} \centering
\includegraphics[width=0.9\columnwidth]{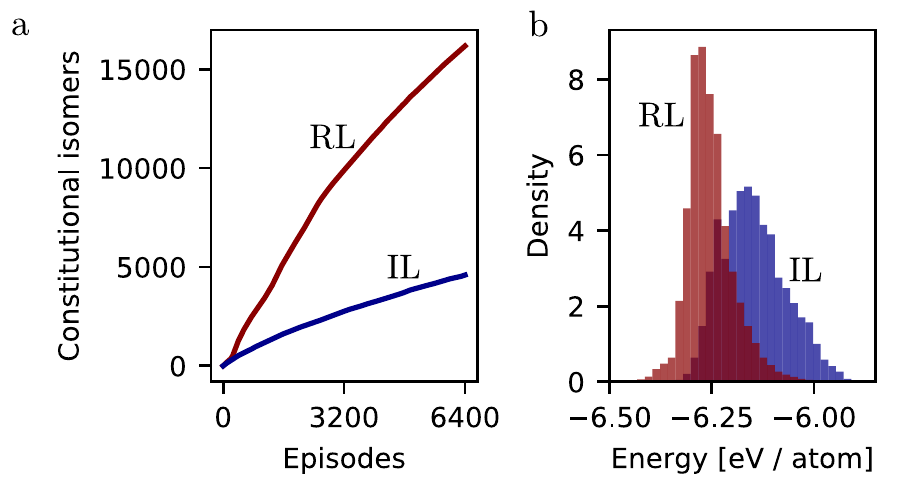}
  \caption{Searching for C$_9$H$_8$O$_4$ isomers. (a) Constitutional
isomers found as the search progresses. As the search progresses the
RL version (red) quickly produces more structures than the IL
version. (b) Energy-density of the generated structures. In addition
to producing more structures, the RL version (red) generates a larger
fraction of low energy molecules.}
  \label{fig:6}
\end{figure}

In Fig.~\ref{fig:6}a the number of unique constitutional isomers
generated as a function of episode number for the IL model
(blue) and the model with RL (red) is seen. Unlike for the smaller
molecules in IV.~C where the IL model was sufficient to cover a large
fraction of configuration space there is now a significant difference
between the RL and IL agent. The RL model clearly outperforms
the IL model in terms of number of molecules generated as the
search progresses, showing that the on-the-fly training is able to
correct inefficiencies in the initial version. However, the
ultimate goal is to generate low energy molecules. Hence, in
Fig.~\ref{fig:6}b the atomization-energy-density of the generated
molecules for both the IL model and the RL version is seen. As
before all stereoisomers are fully relaxed using DFT. The molecules
generated by the RL-enhanced model are significantly more stable than
the IL model, with an average energy difference of 0.105
eV/atom or 2.22 eV per structure. Thus, the RL version does not only
cover more of configuration space, it is also able to focus on the low
energy regions.
\begin{figure}[t]
  \centering
\includegraphics[width=0.95\columnwidth]{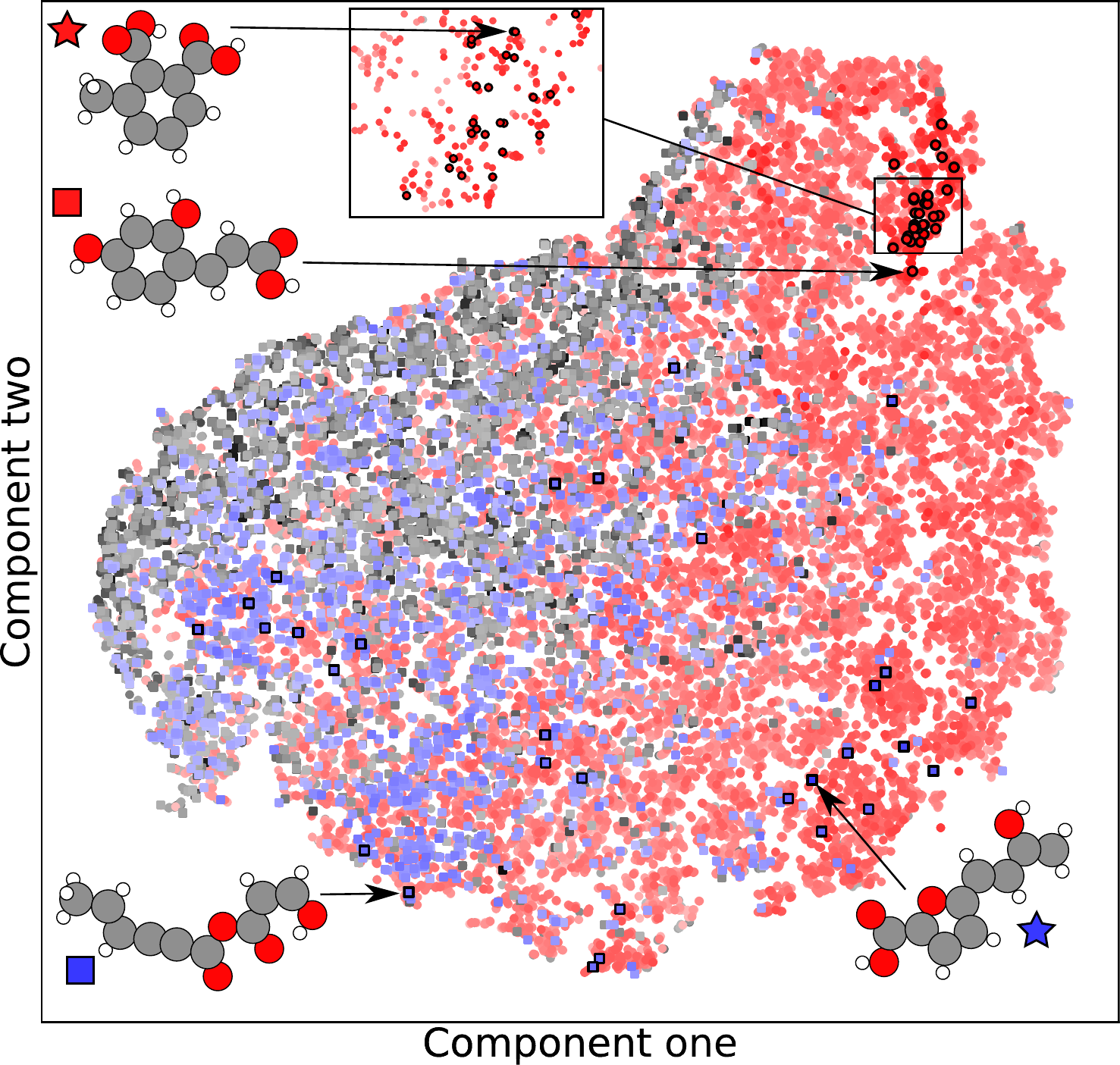}
  \caption{Searching for C$_9$H$_8$O$_4$ isomers. 2D visualization of
IL agent (squares) and RL version (circles). All structures
are colored by their energy on a scale from grey to blue (IL)
or red (RL). The 30 lowest energy structures for both versions of the
algorithms are framed in black.}
  \label{fig:7}
\end{figure}
In Fig.~\ref{fig:7} we investigate a 2D visualization of the generated
molecules for both the IL model (squares colored grey to blue) and RL
version (circles colored grey to red). The structures are represented
using smooth overlap of atomic positions \cite{SOAP} (SOAP) in a 2D
space using t-SNE and colored by their energy using two different
color schemes to help guide the eye. The 30 lowest energy structures
found for both models are framed in black and can be seen in the
supplementary. The RL version discovers certain low energy regions
such as compact aromatic molecules as shown in the inset. The IL
version, however, is not able to extrapolate into this region based on
what it has learned from the database. Interesting molecules among the
low energy isomers include the lowest energy isomer found by the
search (red star) as well as the second lowest isomer (see
supplementary) that both resemble uvitic acid. The search discovers
umbellic acid (red square) and caffeic acid (just outside top
30). Additionally, the RL model discovers a stable isomer with a
7-membered ring (supplementary), a substructure not present in the
original database as only molecules with up to 6 heavy atoms were
used. Despite this, the RL model learns to construct such molecules
which would have been hard to discover by a pure supervised generative
approach. \\
For the IL model we observe fewer of the common aromatic molecules as
it has no driving force towards low energy regions of configuration
space, as exemplified by the lowest energy isomer found (blue star)
being 2 eV higher in energy than the corresponding lowest energy
structure in the RL version. The missing focus on low energy parts of
configuration space results in the 30 lowest energy structures being
scattered across the 2D space, mostly composed of elongated structures
(such as the blue square) where subparts, unlike large aromatic rings,
are more dominant in the GDB-11 database. Utilizing the feedback from
the RL is thus crucial to correct flaws in the IL model as well as
expanding beyond knowledge in the database.

\section{Conclusion}
We have presented a framework for autonomous construction of
molecules. The method relies on a preexisting database which via
supervised learning provides a base level for a model used for
constructing new molecules. Further training of the model is done in a
reinforcement learning setting where feedback is provided by single
point energy calculations by a high-level quantum mechanical total
energy method. The resulting model is able to reproduce structures in
the database as well as producing novel structures. The introduced
model is able to operate directly in 3D space allowing for biasing the
search towards stable molecules. When applying the imitation learning
model in domains far from the training set, the reinforcement learning
procedure corrects initial shortcomings in the model. Further work in
this direction could investigate building molecules in specific
environments, such as organic light-emitting diode (OLED) or organic
solar panels where properties such as HOMO-LUMO gap could be included
in the reward function together with the stability.


\section{Acknowledgments}
We acknowledge support from VILLUM FONDEN (Investigator grant, Project No.\ 16562).
This work has been supported by the Danish National Research
Foundation through the Center of Excellence “InterCat” (Grant
agreement no.: DNRF150)

\section{References}
\bibliography{refs}

\clearpage
\setcounter{figure}{0}
\onecolumngrid
\section{Supplementary}
\subsection{Network details}
In table.~\ref{table:s1} a detailed overview of the blocks in the network
is given. The activation function is a shifted softplus \cite{schnet} (ssp) given by
\begin{equation}
  \text{ssp}(x) = \ln(0.5e^x + 0.5)
\end{equation}

\begin{table}[h]
  \centering
\begin{tabular}{l c c}
  Bag representation & $W_B$ & (5, 16, spp, 32) \\
  Atomwise layers & $W_a$ & (64, 64, ssp) \\
  Distances representation & $W_d$ & (20, 64, ssp) \\
  State Value & $W_{SA}$ & (96, 32, ssp, 1) \\
  Advantage & $W_A$ & (96, 32, ssp, 6) \\
  Energy & $W_E$ & (64, 32, ssp, 16, ssp, 8, ssp, 1) \\
  Concatenation layer & $W_c$ & (64, 64)
\end{tabular}
\caption{Dimension of data as it passes through the fully connected layers. The activation function is is the shifted softplus indicated by ssp.}
\label{table:s1}
\end{table}

\clearpage
\begin{figure*}[h]
  \centering
\includegraphics[width=0.95\textwidth]{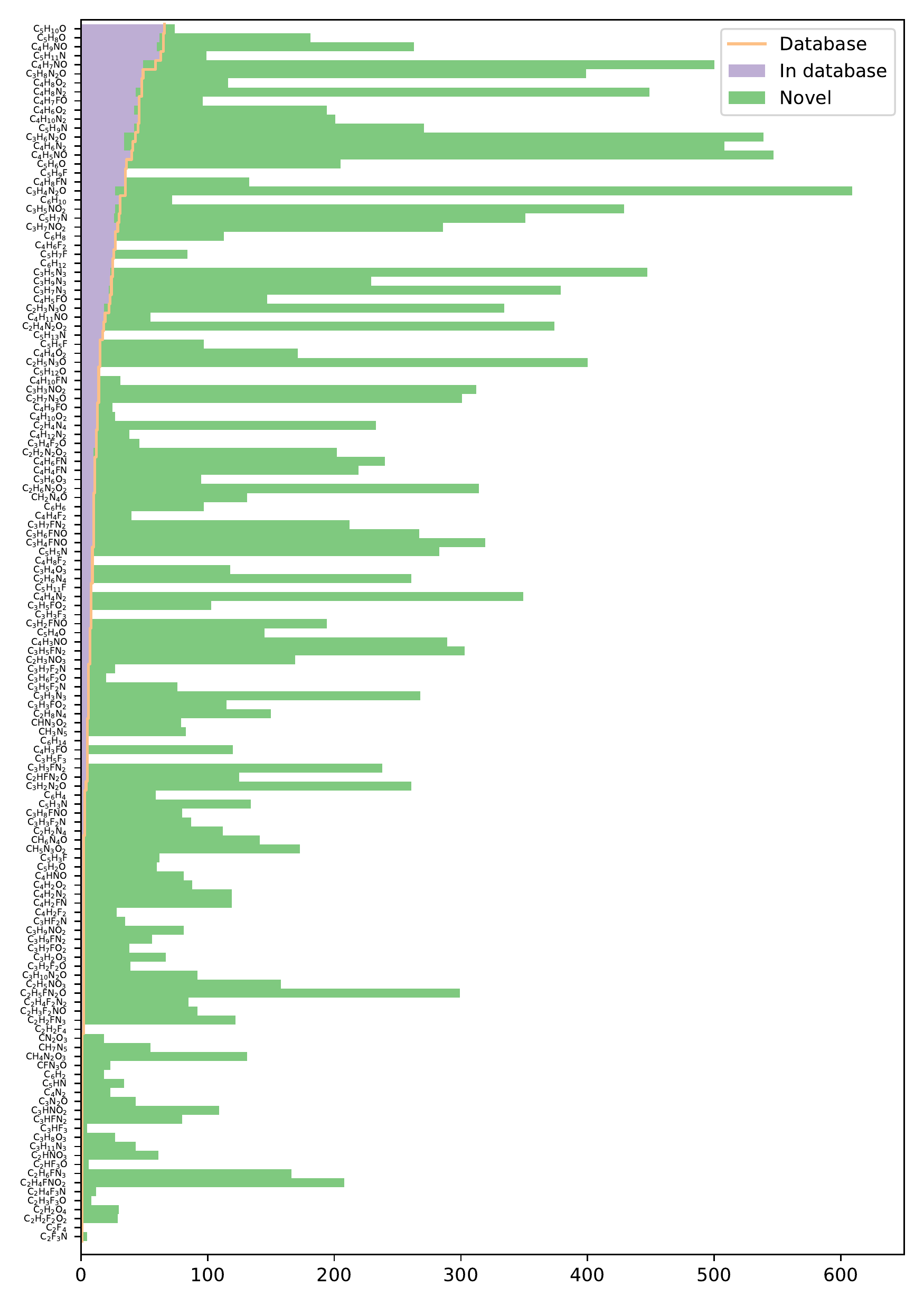}
  \caption{Overview of all constitutional isomers built. Orange line
indicates number of isomers in the database. Purple are isomers built by
ASLA which are also part of the database. Green structures are new
isomers.}
  \label{fig:s1}
\end{figure*}

\clearpage

\begin{figure*}
  \centering
\includegraphics[width=0.90\textwidth]{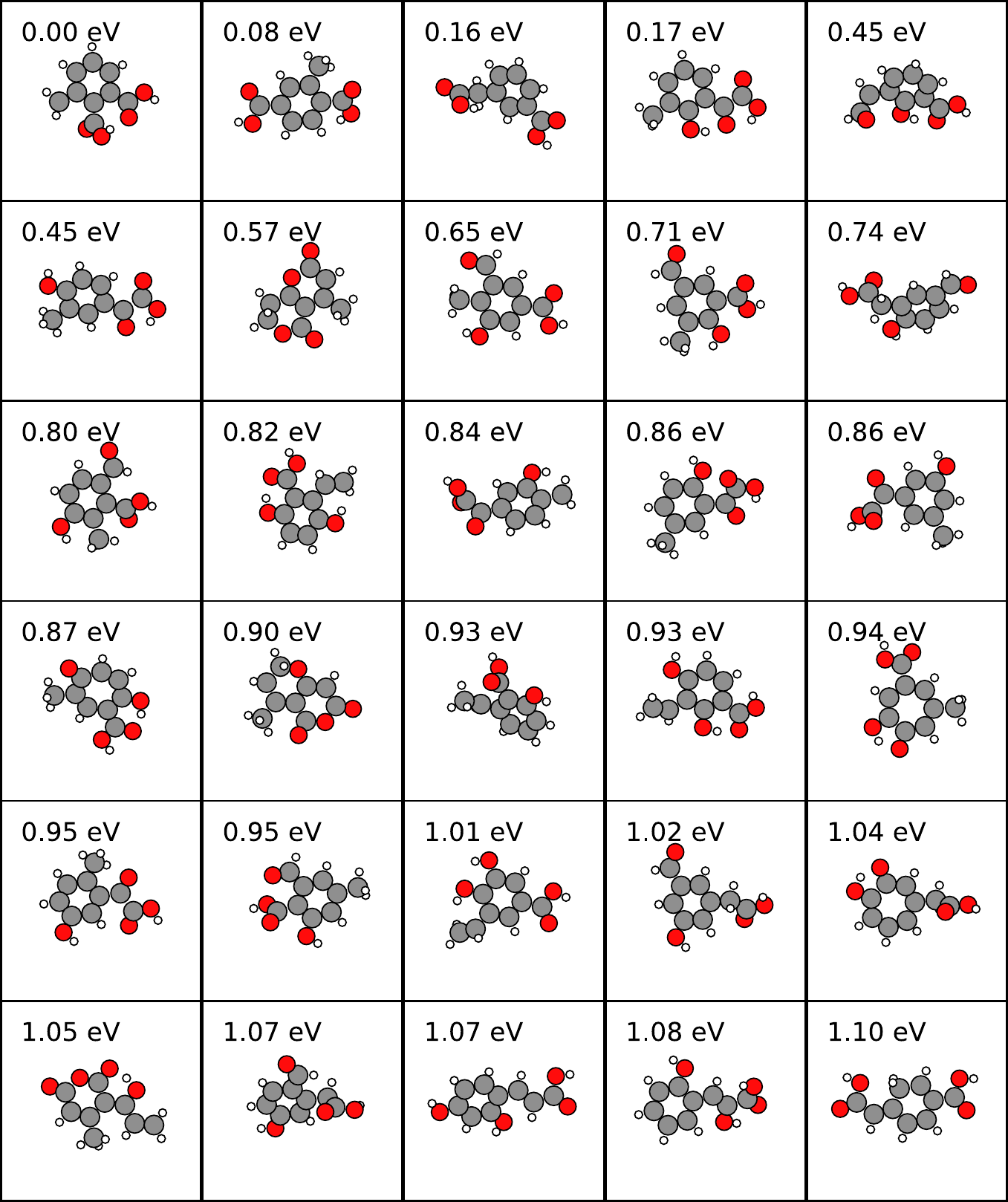}
  \caption{30 lowest energy structures for the RL run.}
  \label{fig:s2}
\end{figure*}

\begin{figure*}[t]
  \centering
\includegraphics[width=0.90\textwidth]{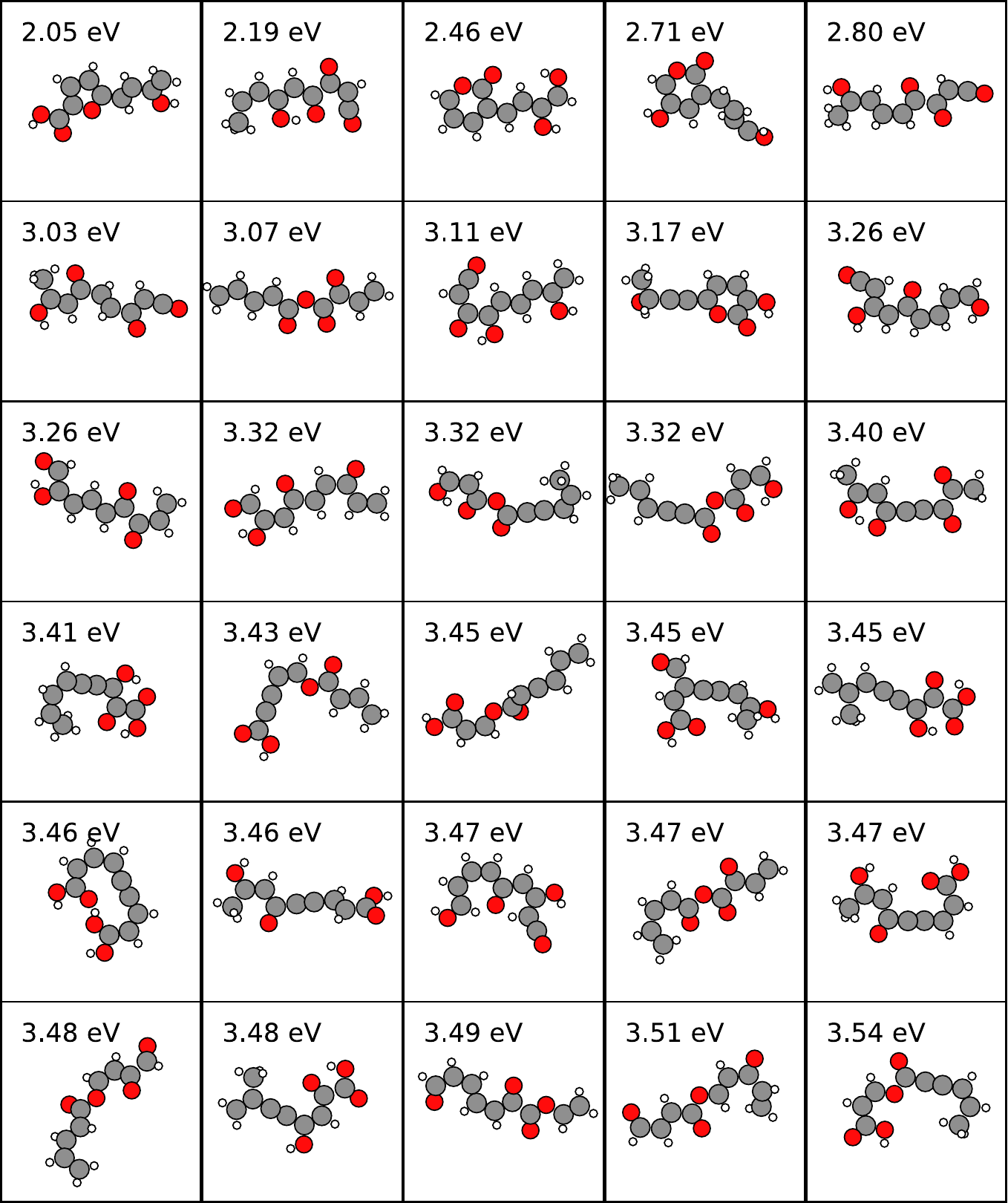}
  \caption{30 lowest energy structures for the IL run. Energies are relative to the lowest energy isomer found in the RL version.}
  \label{fig:s3}
\end{figure*}



\end{document}